\documentclass{article}
\usepackage[margin=1.25in,marginratio=1:1]{geometry}

\usepackage{mathptmx} 
\usepackage{url}

\usepackage{amsmath, amssymb, amsthm}
\usepackage{xcolor, soul}
\usepackage{kbordermatrix}
\usepackage{tikz}
\usetikzlibrary{positioning} 

\usepackage[boxed, noend]{algorithm2e}
\SetAlgoCaptionSeparator{}

\usepackage{mdframed, enumitem}

\theoremstyle{definition}
\newtheorem{example}{Example}[section]
\newmdtheoremenv[innertopmargin=-2pt]{definition}[example]{Definition}
\newmdtheoremenv[skipabove=0pt,innertopmargin=-2pt]{definition-}[example]{Definition}

\newtheorem{manualtheoreminner}{Example}
\newenvironment{manualex}[1]{%
  \renewcommand\themanualtheoreminner{#1}%
  \manualtheoreminner
}{\endmanualtheoreminner}

\newtheoremstyle{dotless}{}{}{}{}{\bfseries}{}{ }{}
 \theoremstyle{dotless}
\newmdtheoremenv[skipabove=0pt, innertopmargin=-2pt]{note}[]{Note:}

\theoremstyle{theorem}
\newmdtheoremenv[innertopmargin=-2pt]{proposition}[example]{Proposition}
\newmdtheoremenv[skipabove=0pt, innertopmargin=-2pt]{proposition-}[example]{Proposition}

\usepackage{caption}
\usepackage{subcaption}

\usepackage{subfiles}

\title{\Large \vspace{-0.6in}An Attempt to Generate Code for Symmetric Tensor Computations\vspace{-0.4in}}
\date{}

\begin{document}
\maketitle
\begin{note}
 	    This document is the culmination of an undergraduate research project conducted by Jessica Shi under the supervision of Stephen Chou, Fred Kjolstad, and Saman Amarasinghe. 
 \end{note}

	\section{Introduction}
	
	Symmetric matrices, a frequently studied topic in linear algebra, can be extended to higher dimensions through \textit{symmetric tensors}, which arise in domains such as computational physics and chemistry \cite{epifanovsky2013new, solomonik2012preliminary}. The fundamental mathematical appeal of the study of symmetry also renders these tensors useful in contexts ranging from a rather beautiful equivalence with homogenous polynomials \cite{comon2008symmetric} to more concrete applications, including decompositions \cite{schatz2014exploiting} and finding eigenvalues \cite{qi2005eigenvalues}.

	Knowing about a tensor's symmetry tells us which of its values are guaranteed to be equal. If leveraged effectively, this can be utilized to improve both storage usage and computation time: values that are the same do not need to be redundantly stored or computed. Many existing works focus on optimizing one specific type of tensor computation involving symmetry.
		
	On the other hand, compilers such as taco \cite{kjolstad2017tensor} can produce code for a broad class of tensor algebra expressions, with support for per-dimension level formats \cite{chou2018format}. However, taco does not presently  provide a way to indicate or apply information about which and how tensors are symmetric. Our goal is thus to work towards a compiler-based approach for symmetric tensor computations, relying on taco for aspects of code generation unrelated to symmetry. 
	 
	 The remainder of this document is outlined as follows. We start in Section 2 by reviewing standard definitions, which characterize a tensor's symmetry by a \textit{partition on its dimensions}, such that coordinates along dimensions that belong to the same part can be permuted without changing the value.
	 
	 Then, we consider in Section 3 the problem of representing symmetric tensors in storage. Some values can be omitted, but when the remaining values are then placed in a dense array, we need new formulas to index into them. While non-symmetric tensors can be geometrically viewed as rectangular prisms, the lower half of, say, a symmetric matrix looks like a triangle. As we move into higher dimensions, the triangles become simplexes, and as a result, our formulas make heavy use of \textit{simplicial numbers}.
	 
	 At this point, we will be able to reason about the symmetries of individual tensors, so we proceed a step further in Section 4 to provide abstractions for thinking about the \textit{symmetry of a computation}, which will ultimately inform the structure of the code that we generate. Interesting cases arise when the various input tensors in the computation are symmetric along dimensions with different indices.
	 
	Section 5 details how to generate a loop structure to handle these cases. Our method is to \textit{separate the for loops} that iterate over a particular dimension based on whether the values in the corresponding region of the input are stored or omitted, in order to determine how to access the tensors correctly. We end by observing that the storage savings exploited in the resulting code come at the cost of poor computational performance.
	 		 	
	\section{Symmetric Definitions}
		
	A matrix $M$ is symmetric if $M[i_1, i_2] = M[i_2, i_1]$. That is, the value is constant under permutations of the indices. We can generalize this definition for tensors \cite{comon2008symmetric}. 
	
	\begin{definition}
		Let $T$ be an $n$-dimensional tensor. Then $T$ is \textit{symmetric} if for all $\pi \in S_n$, $$T[i_1, \dots, i_n] = T[i_{\pi(1)}, \dots, i_{\pi(n)}].$$
	\end{definition}
	
	Here $S_n$ is borrowed from the notation for a symmetric group: for a natural number $n$, let $S_n$ be the set of permutations of $\{1, \dots, n\}$. More generally, for a set $A$, let $S_A$ be the set of permutations of $A$.
		
	\begin{example}
		A three-dimensional tensor $T$ is symmetric if $$T[i_1, i_2, i_3] = T[i_1, i_3, i_2] = T[i_2, i_1, i_3] = T[i_2, i_3, i_1] = T[i_3, i_1, i_2] = T[i_3, i_2, i_1].$$
	\end{example}
	
	The matrix case was restricted to a binary conception of symmetry --- either a matrix is symmetric, or it is not. For higher-order tensors, we can relax the definition. Recall that a \textit{partition} of a set $A$ is a collection of non-empty, pairwise disjoint sets (which we call \textit{parts}) whose union is $A$. Partial symmetries can be defined relative to a chosen partition \cite{schatz2014exploiting}.
	
	\begin{definition}\label{sym}
		Let $T$ be an $n$-dimensional tensor, and let $\{A_i\}_{i = 1}^m$ be a partition of the tensor dimensions $\{1, \dots, n\}$. Then $T$ is \textit{partially symmetric} if for all $(\pi_1, \dots, \pi_m) \in S_{A_1} \times \dots \times S_{A_m}$, $$T[i_1, \dots, i_n] = T[i_1', \dots, i_n'],$$ where $$i_j' = \begin{cases} i_{\pi_1(j)} & \text{if } j \in A_1 \\ \hfill \vdots \hfill \\  i_{\pi_m(j)} & \text{if } j \in A_m \end{cases}.$$
		In particular, we say $T$ has $\{A_i\}_{i = 1}^m$ symmetry. 
	\end{definition}
	
	\begin{example}\label{partial}
		A four-dimensional tensor $T$ has $\{{\color{teal} \{1, 3\}}, {\color{magenta}\{2, 4\}}\}$ symmetry if $$T[i_1, i_2, i_3, i_4] = T[{\color{teal} i_3}, i_2, {\color{teal} i_1}, i_4] = T[i_1, {\color{magenta} i_4}, i_3, {\color{magenta} i_2}] = T[{\color{teal} i_3}, {\color{magenta} i_4}, {\color{teal} i_1},  {\color{magenta} i_2}].$$ 
	\end{example}
	
	Observe that ``partial'' symmetry is a bit of a misnomer because it encompasses the other definitions. If $T$ is symmetric, or more precisely, \textit{fully symmetric}, then  the partition has all dimensions in one part: $\{\{1, \dots, n\}\}.$ If $T$ is \textit{non-symmetric}, then the partition has each dimension in its own part: $\{\{i\}\}_{i = 1}^n.$

	Though partitioning on the dimensions allows us to define partial symmetry independently of the names of the index variables, this can become notationally burdensome. So, we will usually express situations such as Example \ref{partial} as a tensor $T[i, j, k, \ell]$ having $\{\{i, k\}, \{j, \ell\}\}$ symmetry.
			
	We can also take a moment to consider some types of symmetries not encompassed by Definition \ref{sym}: this might include anti-symmetry or more broadly, symmetries where values are equal up to some transformation; symmetries where blocks, rather than individual elements, are equal; or symmetries that hold under a more limited group of permutations --- only when the indices are cycled, for example. 
		
	\section{Symmetric Storage}
	
	By definition, tensors that have symmetry contain repeated values, and in the previous section, we defined a systematic way to notate these redundancies. In this section, we establish a storage scheme that only maintains one coordinate-value pair for each set of coordinates that are guaranteed to have the same value.
		
	\subsection{Approach}
	
	We again start with the simplest case: a symmetric matrix.
	
	\begin{example}\label{symstor}
		Suppose we have $$M = \begin{bmatrix} 1 & 2 & 4 \\ 2 & 3 & 5 \\ 4 & 5 & 6 \end{bmatrix}.$$ We only need to store values in, say, the lower half. So, we can represent $M$ as the dense array $[1, 2, 3, 4, 5, 6].$
	\end{example}
		
	The choice of which half to store was mostly arbitrary, but we should remain consistent. We will deem some coordinates of a symmetric tensor to be ``canonical,'' which is a term borrowed from \cite{epifanovsky2013new}. 
			
	\begin{definition}
		Let tensor $T[a_1, \dots, a_n]$ have symmetry $S$. Coordinates $[c_1, \dots, c_n]$ are \textit{canonical} if $c_i \geq c_j$ for any $i < j$ with $a_i$ and $a_j$ in the same part of $S$. Otherwise, the coordinates are \textit{non-canonical}.
	\end{definition}
	
	\begin{example}
		For a symmetric matrix $M[i, j]$, canonical coordinates have $i \geq j$, so they are in the lower half.
	\end{example}
	
	\begin{example}
		For a tensor $T[i, j, k]$ with symmetry $\{\{i\}, \{j, k\}\}$, canonical coordinates have $j \geq k$. So, $[3, 2, 1]$ and $[1, 3, 2]$ would be canonical, but $[2, 1, 3]$ would not.
	\end{example}
	
	Though there are varied approaches to symmetric storage, we use one that eliminates all redundancy. If a tensor is \textit{stored symmetrically}, then we only store values for canonical coordinates, as in Example \ref{symstor}. Naturally, it is possible for symmetric tensors not to be stored symmetrically, but for simplicity, we will assume that they are unless explicitly noted. We rephrase and reiterate the invariant for our storage scheme: $$\text{Only the canonical coordinates of a symmetric tensor may be accessed.}$$
	Two questions follow. How do we access the canonical coordinates whose values we want to retrieve or set (Section 3.2)? And what happens if we want to access non-canonical coordinates (Section 3.3)?
	
	\subsection{Triangular Accesses}\label{tri}
		
	We need formulas to randomly access symmetric tensors. In this section, let $N_i$ denote the size of the dimension indexed by $i$. We start by comparing the formulas for a non-symmetric versus symmetric matrix. 
	
	\begin{example}
		For a non-symmetric matrix whose values are stored in a dense array, the element at $(i, j)$ is located at position $i \cdot N_i + j$ of the array. The first term represents the position of $(i, 0)$, since $$\underbrace{N_i + N_i + \dots + N_i}_{i \text{ times}} = i \cdot N_i,$$ and the second term represents the offset needed for the remaining $(0, j)$.
	\end{example}
	
	\begin{example}\label{ind2}
		For a symmetric matrix, the position of $(i, 0)$ needs to be computed differently, since there is just one canonical value in the first row, two in the second, and so on. This gives $$1 + 2 + \dots + i = \frac{i \cdot (i + 1)}{2} = t_i.$$
		Here, $t$ stands for \textit{triangular number}. The offset for $j$ remains the same. Hence the position of $(i, j)$ is $t_i + j$.
	\end{example}
	
	Next, consider what happens upon advancing to three dimensions.
	
	\begin{example}\label{ind3}
		For a fully symmetric tensor, to find the position of $(i, j, k)$, we again start by computing the position of $(i, 0, 0)$. Implicitly, we count the number of canonical coordinates that appear earlier in the tensor. Recall that canonical means the coordinates are in non-increasing order.
		
		Fix a first coordinate $i'$. Then, fix a second coordinate $j'$. The number of canonical possibilities for a third coordinate is $j' + 1$. Summing over $j'$ such that $0 \leq j' \leq i'$, the number of indices starting with $i'$ is $$1 + 2 + \dots + (i' + 1) = t_{i' + 1}.$$ Triangular numbers appear again! Summing over $i'$ such that $0 \leq i' < i$, we have that $t_1 + \dots + t_i$ is the position of $(i, 0, 0)$. We interrupt the example with a mathematical definition of these types of summations \cite{polytopic}.
	\end{example}
	
	\begin{definition-}
		The $n$-th $d$-simplicial number is $s_{d}(n)$. The base case is $s_0(n) = 1$. For $d > 0$, the recurrence relation is $$s_{d}(n) = \sum_{i = 1}^{n} s_{d - 1}(i).$$ 
	\end{definition-}
	
	The terminology has geometrical origins: a \textit{simplex} is a generalization of the triangle for arbitrary dimensions. When $d = 1$, we simply have the natural numbers. When $d = 2$, we have sums of natural numbers, called triangular numbers. When $d = 3$, we have sums of triangular numbers, called \textit{tetrahedral numbers}. 
		
	\begin{proposition}
		The closed form is given by $$s_d(n) = \frac{1}{d!} \cdot \prod_{i = 1}^{d} (n + d - i).$$
	\end{proposition}
	
	We return to our previous examples.
	
	\begin{manualex}{\ref{ind2}}[Revisited]
		We can rewrite the formula $t_i + j$ in terms of simplicial numbers: $s_2(i) + s_1(j)$. 
	\end{manualex}
	
	\begin{manualex}{\ref{ind3}}[Revisited]
		The position of $(i, 0, 0)$ is $s_3(i).$ The slice of the tensor at the chosen $i$ is just a symmetric matrix over indices $(j, k)$, a case we already have a formula for. So, the position of $(i, j, k)$ is $$s_3(i) + s_2(j) + s_1(k).$$
	\end{manualex}
	
	In general, for an $n$-dimensional symmetric tensor, we can reduce to the $(n-1)$-dimensional case.
	
	\begin{proposition}
		The element at $(i_1, \dots, i_n)$ in an $n$-dimensional fully symmetric tensor is located at position $\sum_{i = 1}^{n} s_i(n)$ of the corresponding dense array.
	\end{proposition}
	
	To extend the formula for partial symmetries, we begin with examples in four and five dimensions.
	
	\begin{example}
		Consider a tensor with $\{\{i, j, k\}, \{\ell\}\}$ symmetry. The position of $(i, j, k, 0)$ is $$\left(s_3(i) + s_2(j) + s_1(k) \right) \cdot s_1(N_\ell).$$
		The parenthetical term represents the possibilities for the first three coordinates, which are constrained by the symmetry to be canonically non-increasing, and the factor of $s_1(N_\ell) = N_\ell$ represents the possibilities for the fourth coordinate, which is not constrained. To find the position of $(i, j, k, \ell)$, we simply add $\ell$. 
	\end{example}
	
	\begin{example}\label{part1}
		Consider a tensor with $\{{\color{orange} \{i, j\}}, {\color{teal} \{k, \ell\}}, {\color{magenta} \{m\}}\}$ symmetry. As we did above, we focus on one part of the partition at a time. The position of $(i, j, 0, 0, 0)$ is $$\left({\color{orange} s_2(i) + s_1(j)} \right) \cdot {\color{teal} s_2(N_k)} \cdot {\color{magenta} s_1(N_m)},$$ where the terms represent the possibilities for the coordinates in the part of the symmetry of the same color. For example, for ${\color{teal} \{k, \ell\}}$, there are $N_k$ possibilities for the third coordinate, since it is unconstrained relative to the previous two, but the fourth coordinate must be at most the third, so this gives the $N_k$-th triangular number, ${\color{teal} s_2(N_k)}$. 
		The position of $(i, j, k, \ell, 0)$ is the above plus $$\left({\color{teal} s_2(k) + s_1(\ell)}\right) \cdot {\color{magenta} s_1(N_m)}.$$ Finally, the position of $(i, j, k, \ell, m)$ is the two terms above plus the additional offset ${\color{magenta} s_1(m)}$. 	\end{example}
		
	We can generalize the above observations into a formula. Notationally, let the tensor have $$\{\{a_{i, j}\}_{j = 1}^{p_i}\}_{i = 1}^n$$ symmetry, such that this set of sets is a partition of the indices. Because we can permute the indices in any part, we must also have for each $i$ that $N_{a_{i, 1}} = \dots = N_{a_{i, p_i}}$.
	
	\begin{proposition}\label{form1}
		The element at ${(a_{i,j})_{j = 1}^{p_i}}_{i = 1}^{n} = (a_{1, 1}, \dots, a_{1, p_1}, \dots, a_{n, p_n})$ is located at position $$\sum_{i = 1}^n\left( \sum_{j = 1}^{p_i} s_{p_i - j + 1}(a_{i, j}) \cdot \prod_{i' = i + 1}^n s_{p_{i'}}(N_{a_{i', 1}}) \right)$$ of the corresponding dense array.  
	\end{proposition}
	
	We can familiarize ourselves with the notation through an example.
	
	\begin{manualex}{\ref{part1}}[Revisited]
		Recall that the tensor has $\{\{i, j\}, \{k, \ell\}, \{m\}\}$ symmetry. There are three parts, so $n = 3$, and the respective sizes of the parts lead to $p_1 = p_2 = 2$ and $p_3 = 1$. Moreover, we have $$a_{1, 1} = i, \ a_{1, 2} = j, \ a_{2, 1} = k, \ a_{2, 2} = \ell, \ a_{3,1} = m.$$
		
		To compute the position of $(i, j, k, \ell, m)$, we need to sum three terms. The $\{i, j\}$ part contributes $$\left(s_{p_1 - 1 + 1} (a_{1, 1}) + s_{p_1 - 2 + 1}(a_{1, 2}) \right) \cdot \left(s_{p_2}(N_{a_{2, 1}}) \cdot s_{p_3}(N_{a_{3,1}})\right) = (s_2(i) + s_1(j)) \cdot s_2(N_k) \cdot s_1(N_m),$$ the $\{k, \ell\}$ part contributes $$\left(s_{p_2 - 1 + 1}(a_{2,1}) + s_{p_2 - 2 + 1}(a_{2,2})\right) \cdot s_{p_3}(N_{a_{3,1}}) = (s_2(k) + s_1(\ell)) \cdot s_1(N_m),$$ and the $\{m\}$ part contributes $$s_{p_3 - 1 + 1}(a_{3,1}) \cdot 1 = s_1(m).$$
		These terms match what was previously found.
	\end{manualex}

	We can also rewrite the formula recursively by factoring out the common products:
	
	\begin{proposition}\label{form2} For $i$, $j$ defined above, let
			$$\mathbb{I}(a_{i, j}) = \begin{cases}
				\mathbb{I}(a_{i - 1, p_{i - 1}}) \cdot s_{p_i} (N_{a_{i, 1}}) + s_{p_i} (a_{i, 1})	   & \text{if } j = 1\\
				\mathbb{I}(a_{i, j - 1}) + s_{p_i - j + 1} (a_{i, j}) & \text{if } j \neq 1		
			\end{cases}.$$
		As a base case, let $\mathbb{I}$ evaluate to $0$ when $i = 0$. The element at ${(a_{i,j})_{j = 1}^{p_i}}_{i = 1}^{n}$ is located at position $\mathbb{I}(a_{n, p_n})$. 
	\end{proposition}
	
	Again, we examine some examples.
	
	\begin{manualex}{\ref{part1}}[Revisited] If we use the above definition, we obtain
		\begin{align*}
			\mathbb{I}(i) &= s_2(i)\\
			\mathbb{I}(j) &= \mathbb{I}(i) + s_1(j)\\
			\mathbb{I}(k) &= \mathbb{I}(j) \cdot s_2(N_k) + s_2(k)\\
			\mathbb{I}(\ell) &= \mathbb{I}(k) + s_1(\ell)\\
			\mathbb{I}(m) &= \mathbb{I}(\ell) \cdot s_1(N_m) + s_1(m).
		\end{align*}
	\end{manualex}
	
	Observe that nothing particularly exciting happened in the process of moving from Proposition \ref{form1} to Proposition \ref{form2} --- we just rearranged the terms. The latter set of formulas, however, are the ones that would appear in the generated code, as we want to reuse common terms rather than recomputing for each new index. Indeed, we can check that this aligns with what we expect to have in the non-symmetric case.
	
	\begin{example}
		For a non-symmetric tensor with indices $i_1, \dots, i_n$, its ``symmetry'' is a partition with  $n$ parts, each of size $p_i = 1$. Then, we have
		\begin{align*}
			\mathbb{I}(i_1) &= s_1(i_1) \\
			\mathbb{I}(i_2) &= \mathbb{I}(i_1) \cdot s_1(N(i_2)) + s_1(i_2)\\
			\vdots\\
			\mathbb{I}(i_n) &= \mathbb{I}(i_{n - 1}) \cdot s_1(N(i_{n})) + s_1(i_n).
		\end{align*}
		
		If we remind ourselves that $s_1(x) = x$ for any $x$, then we can see that we have arrived at our usual formulas for indexing into dense, non-symmetric tensors.
	\end{example}
	
	A major limitation of these formulas lurks in the phrase ``corresponding dense array.''  Propositions \ref{form1} and \ref{form2} only hold if the dimensions of the tensors are stored in the same order as they appear in the symmetry. For example, if we have a tensor $T[i, j, k]$ with $\{\{i, k\}, \{j\}\}$ symmetry, we know the formula in the case where the tensor is stored in $i \rightarrow k \rightarrow j$ order, but not in the standard (row-major) $i \rightarrow j \rightarrow k$ order. It appears theoretically plausible that general formulas for these cases exist, but we were not able to find them.
	
	\subsection{Loop Structure (Example)}\label{loopex}	
	
	\begin{figure}[t]
		\vspace{-\baselineskip}
		\begin{subfigure}[b]{0.48\textwidth}
			\begin{subfigure}[b]{\textwidth}
			\begin{algorithm}[H]
				\For{$i$}{
					\For{$j$}{
						$C[i, j] = A[i, j] + B[i, j]$
					}
				}
			\end{algorithm}
			\caption{Both non-symmetric}
			\label{loops:a}
			\end{subfigure}
			
			\vspace{1em}
		
			\begin{subfigure}[b]{\textwidth}
			\begin{algorithm}[H]
				\For{$i$}{
        					\For{${\color{magenta} j \leq i}$}{
            					$C[i, j] = A[i, j] + B[i, j]$
        					}
        				}
			\end{algorithm}
			\caption{Both symmetric}
			\label{loops:b}
			\end{subfigure}
		\end{subfigure}
		\hfill
		\begin{subfigure}[b]{0.48\textwidth}
			\begin{algorithm}[H]
				\For{$i$}{
        					\For{$j \leq i$}{
            					$C[i, j] = A[i, j] + B[i, j]$
        					}				
					\For{{\color{magenta} $j > i$}}{
						$C[i, j] = {\color{magenta} A[j, i]} + B[i, j]$
					}
        				}
			\end{algorithm}
			\caption{Only $A$ symmetric}
			\label{loops:c}
		\end{subfigure}
		\caption{Possible algorithms for matrix addition, depending on the symmetries of the inputs.}
		\label{loops}
	\end{figure}

	In this section, we focus on the matrix addition $$C[i, j] = A[i, j] + B[i,j]$$ and its accompanying pseudocode algorithm in Figure \ref{loops:a}.
	
	If the inputs $A$ and $B$ are both symmetric, then $C[i, j] = A[j, i] + B[j, i] = C[j, i]$, so the output $C$ is symmetric also. We only need to compute and store the values for the canonical half of coordinates in $C$, which are coordinates such that $j \leq i$. This results in the modification to the loop bound shown in Figure \ref{loops:b}.

	If $A$ is symmetric and $B$ is not, observe that $C$ is not symmetric, so we must compute both when $j \leq i$ and when $j > i$. For the former, the loop with this condition is the same as before. For the latter, a problem arises: $A$ is symmetric, so the access $A[i, j]$ is non-canonical and thus invalid when $j > i$. But precisely due to $A$'s symmetry, we can convert to the canonical access $A[j, i]$. Hence we obtain the algorithm in Figure \ref{loops:c}.
		
	One way we can describe what has happened here is that when $A$ and $B$ are both symmetric, we preserve a similar loop structure as the non-symmetric case. When exactly one of $A$ and $B$ is symmetric, however, the loop structure looks different because we need to traverse canonical \textit{and} non-canonical regions of the input. The goal of the next two sections is to develop a method to determine, for a more general class of computations, when and what changes need to be made.

	\section{Symmetric Computations}
	
	\subsection{Approach}		
 	How do we compute with symmetric tensors? Our assumption will be that we are provided a tensor computation and the \textit{input symmetries} of the tensors on the right-hand side of the equation.
	
	Notice we are implicitly claiming we can figure out the \textit{output symmetry} of the tensor on the left-hand side, instead of requiring that this information be manually supplied. In the matrix addition example, the output was symmetric if and only if both inputs were symmetric. Is it always this simple? Naturally not.
	
	\begin{example}\label{matmul}
		For the matrix multiplication $$C[i, k] = A[i, j] * B[j, k],$$ even if $A$ and $B$ are both symmetric, $C$ will not be symmetric unless $AB = BA$.
	\end{example}
	
	So, we will need to derive the output symmetry in a way that takes into account the indices used in the input symmetries. It also does not suffice to only examine the output: 

	\begin{example}\label{contract}
		Consider a contraction such as $$C[i, \ell] = A[i, j, k] * B[j, k, \ell].$$
		If $A$ has $\{\{i\}, \{j, k\}\}$ symmetry, its canonical coordinates have $j \geq k$. If $B$ is non-symmetric, we also need to access coordinates with $j < k$. Though $j$ and $k$ do not appear in the output, they can  affect the loop structure.
	\end{example}
	
	In other words, we originally defined symmetry as a property of individual tensors, but we can now also define the symmetry of a computation, which we give a name to below:
		
	\begin{definition}\label{gcs}
		Given a computation and its input symmetries, the \textit{greatest common symmetry} (gcs) is a partition of the set of all indices in the computation. Indices $i_1$ and $i_2$ are in the same part of the gcs if for every tensor $T$ in the computation, one of the following is true:
		\begin{itemize}[noitemsep, topsep=0.5em]	
			\item $i_1$ and $i_2$ are indices of $T$, and $T$ is the output tensor; or
			\item $i_1$ and $i_2$ are indices of $T$, and they belong to the same part of $T$'s symmetry; or
			\item $i_1$ and $i_2$ are not indices of $T$.
		\end{itemize}
	\end{definition}
		
	\begin{manualex}{\ref{matmul}}[Revisited]
		For each pair of indices, there is a tensor such that one index is present and the other is not. Hence regardless of the input symmetries, the gcs of this matrix multiplication is $\{\{i\}, \{j\}, \{k\}\}$.
	\end{manualex}
	
	\begin{manualex}{\ref{contract}}[Revisited]
		For indices $j$ and $k$, they are in the same part of the gcs if and only if they are in the same part of both $A$ and $B$'s symmetries. In this case, the gcs is  $\{\{i\}, \{j, k\}, \{\ell\}\}$.
	\end{manualex}
	
	Earlier, we claimed that we could derive the output symmetry of a computation. This is a two part process. First, we use the input symmetries to compute the gcs. The second step is described below.
	\begin{definition}\label{incl}
		Let $S_1$ and $S_2$ be symmetries. Then $S_1 / S_2$ is a partition of the indices of $S_1$ such that $i_1$ and $i_2$ are in the same part if they are in the same part of $S_1$ and $S_2$.
	\end{definition}
	
	Observe that $S_1 / S_2$ is a refinement of $S_1$: that is, every part in $S_1/S_2$ is a subset of a part in $S_1$.
	
	\begin{proposition}\label{out}
		Let $S_G$ be the gcs of some computation, and let $O$ be the set of indices in the output tensor. The output symmetry is $\{O\} / S_G$.
	\end{proposition}

	 \begin{figure}
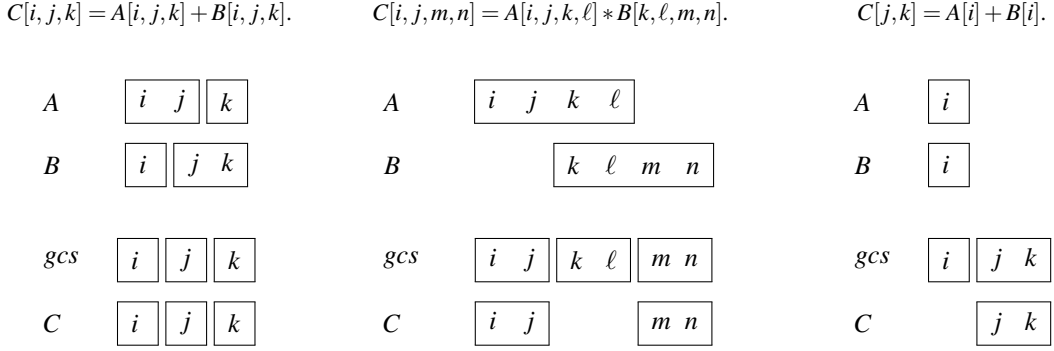

	 	\vspace{-\baselineskip}
		\begin{subfigure}[b]{0.3\textwidth}
	 	\centering
		\caption*{$C[i, j, k] = A[i, j, k] + B[i, j, k]$.}
		\ \\
		\ \\
		\setlength{\fboxsep}{0.5em}
	 	\begin{minipage}{0.29\textwidth}\hspace{1em} $A$ \end{minipage}
		\begin{minipage}{0.5\textwidth}
		\hspace{0.05em}
		\fbox{\begin{minipage}[t][0.6em][c]{1.75em}{
		 $i$ \hfill $j$}
		 \end{minipage}} 
		 \fbox{\begin{minipage}[t][0.6em][c]{0.5em}{
		 $k$}
		 \end{minipage}} 
		\end{minipage}
		
		\ \\
		
		\begin{minipage}{0.29\textwidth}\hspace{1em} $B$ \end{minipage}
		\begin{minipage}{0.5\textwidth}
		\hspace{0.05em}
		\fbox{\begin{minipage}[t][0.6em][c]{0.5em}{
		 $i$}
		 \end{minipage}} 
		 \fbox{\begin{minipage}[t][0.6em][c]{1.75em}{
		 $j$ \hfill $k$}
		 \end{minipage}} \end{minipage}
		
		\ \\
		\ \\
		
		\begin{minipage}{0.29\textwidth}\hspace{1em} $gcs$ \end{minipage}
		\begin{minipage}{0.5\textwidth}
		\fbox{\begin{minipage}[t][0.6em][c]{0.5em}{
		 $i$} \end{minipage}} 
		 \fbox{\begin{minipage}[t][0.6em][c]{0.5em}{
		 $j$}  \end{minipage}} 
		 \fbox{\begin{minipage}[t][0.6em][c]{0.5em}{
		 $k$}  \end{minipage}} 
		 \end{minipage}
		
		\ \\
		
		\begin{minipage}{0.29\textwidth}\hspace{1em} $C$ \end{minipage}
		\begin{minipage}{0.5\textwidth}
		\fbox{\begin{minipage}[t][0.6em][c]{0.5em}{
		 $i$}  \end{minipage}} 
		 \fbox{\begin{minipage}[t][0.6em][c]{0.5em}{
		 $j$}  \end{minipage}} 
		 \fbox{\begin{minipage}[t][0.6em][c]{0.5em}{
		 $k$} \end{minipage}} 
		 \end{minipage}
		\end{subfigure}
		\hfill
		\begin{subfigure}[b]{0.35\textwidth}
	 	\centering
		\caption*{$C[i, j, m, n] = A[i, j, k, \ell] * B[k, \ell, m, n]$.}
		\ \\
		\ \\
		\setlength{\fboxsep}{0.5em}
	 	\begin{minipage}{0.29\textwidth}\hspace{1em} $A$ \end{minipage}
		\begin{minipage}{0.69\textwidth}
		\fbox{\begin{minipage}[t][0.6em][c]{5em}{
		$i$ \hfill $j$ \hfill $k$ \hfill $\ell$}
		\end{minipage}} \end{minipage}
		
		\ \\
		
		\begin{minipage}{0.29\textwidth}\hspace{1em} $B$ \end{minipage}
		\begin{minipage}{0.69\textwidth}  \hspace{2.75em}
		\fbox{\begin{minipage}[t][0.6em][c]{5em}{
		$k$ \hfill $\ell$ \hfill $m$ \hfill $n$}
		\end{minipage}} \end{minipage}
		
		\ \\
		\ \\
		
		\begin{minipage}{0.29\textwidth}\hspace{1em} $gcs$ \end{minipage}
		\hfill 
		\begin{minipage}{0.69\textwidth}
		\fbox{\begin{minipage}[t][0.6em][c]{1.75em}{
		 $i$ \hfill $j$}
		 \end{minipage}} 
		 \fbox{\begin{minipage}[t][0.6em][c]{1.75em}{
		   $k$ \hfill $\ell$}
		 \end{minipage}} 
		  \fbox{\begin{minipage}[t][0.6em][c]{1.75em}{
		  $m$ \hfill $n$}
		 \end{minipage}} 
		 \end{minipage}
		
		\ \\
		
		\begin{minipage}{0.29\textwidth}\hspace{1em} $C$ \end{minipage}
		\hfill 
		\begin{minipage}{0.69\textwidth}
		\fbox{\begin{minipage}[t][0.6em][c]{1.75em}{
		 $i$ \hfill $j$}
		 \end{minipage}} 
		 \hphantom {\fbox{\begin{minipage}[t][0.6em][c]{1.75em}{
		   $k$ \hfill $\ell$}
		 \end{minipage}}}
		  \fbox{\begin{minipage}[t][0.6em][c]{1.75em}{
		  $m$ \hfill $n$}
		 \end{minipage}} 
		 \end{minipage}
		\end{subfigure}
		\hfill
		\begin{subfigure}[b]{0.3\textwidth}
	 	\centering
		\caption*{$C[j, k] = A[i] + B[i]$.}
		\ \\
		\ \\
		\setlength{\fboxsep}{0.5em}
	 	\begin{minipage}{0.29\textwidth}\hspace{1em} $A$ \end{minipage}
		\begin{minipage}{0.45\textwidth}
		\fbox{\begin{minipage}[t][0.6em][c]{0.5em}{
		 $i$} \end{minipage}} 
		 \end{minipage}
		
		\ \\
		
		\begin{minipage}{0.29\textwidth}\hspace{1em} $B$ \end{minipage}
		\begin{minipage}{0.45\textwidth}
		\fbox{\begin{minipage}[t][0.6em][c]{0.5em}{
		 $i$} \end{minipage}}  \end{minipage}
		
		\ \\
		\ \\
		
		\begin{minipage}{0.29\textwidth}\hspace{1em} $gcs$ \end{minipage}
		\begin{minipage}{0.45\textwidth}
		\fbox{\begin{minipage}[t][0.6em][c]{0.5em}{
		 $i$}
		 \end{minipage}} 
		 \fbox{\begin{minipage}[t][0.6em][c]{1.75em}{
		 $j$ \hfill $k$}
		 \end{minipage}} 
		 \end{minipage}
		
		\ \\
		
		\begin{minipage}{0.29\textwidth}\hspace{1em} $C$ \end{minipage}
		\begin{minipage}{0.45\textwidth}
		\hphantom{\fbox{\begin{minipage}[t][0.6em][c]{0.5em}{
		 $i$}
		 \end{minipage}}}
		 \fbox{\begin{minipage}[t][0.6em][c]{1.75em}{
		 $j$ \hfill $k$}
		 \end{minipage}} 
		 \end{minipage}
		\end{subfigure}
		
		\caption{Examples show, for a computation and its input symmetries, what the greatest common symmetry (gcs) and output symmetry look like. Indices in the same box are in the same part of the partition.}
		\label{boxes}
	 \end{figure}

	Several examples can be found in Figure \ref{boxes}. To build intuition about the claim, consider a computation $L = ?_{i = 1}^n R_i,$ where $L$ is the output tensor and the $R_i$ are the input tensors. We use $?$ to represent sums or products. Let $S_O$ be the output symmetry determined using Proposition \ref{out}. 
	
	To show that $L$ does have $S_O$ symmetry, consider any permutation $\pi$ allowed by $S_O$. Apply $\pi$ to the indices of $L$ to obtain $L'$ and the indices of the $R_i$ to obtain $R_i'$. Due to our definitions above, we know that the value of each $R_i$ must be constant under $\pi$. So, $$L' =  ?_{i = 1}^n R_i' = ?_{i = 1}^n R_i = L.$$ The value of $L$ is also constant under $\pi$, as desired.

	How far have we progressed towards our goal of determining the structure of symmetric computations? For any part in the gcs, we should access those indices only in accordance with the canonical ordering. If parts of the input symmetries are strict supersets of parts in the gcs, this will suggest that we need to change the loops in order to avoid non-canonical accesses. We concretize these ideas in Section 5.
	
	\subsection{Limitations}
	
	Our methods focus on the indices of the tensors rather than the tensors themselves, so we fail to find certain symmetries. As mentioned in Example \ref{matmul}, matrix multiplication will produce a symmetric output in the case where the input matrices commute, which is outside of the scope of the gcs. 
	
	Additionally, if the same tensor appears multiple times, the computation can be inherently symmetric. Consider the change-of-basis operation described in \cite{schatz2014exploiting}. In the matrix case, we have $C = X A X^T.$ Transposing, we have $C^T = X A^T X^T,$ so $C$ is symmetric when $A$ is symmetric. But when this computation is rewritten in our notation as $$C[i, \ell] = X[i, j] * A[j, k] * X[\ell, k],$$ this relationship is no longer detectable under our definitions.
		 
	\section{Symmetric Code}
	
	Now that we are equipped with a way to describe the symmetries of both individual tensors and computations, we can develop an approach and algorithm for symmetric code generation. These will be centered around the idea that in order for the code to be simple enough that it is easily generalizable, we need to make the relationships between non-canonical coordinates and their canonical counterparts apparent. 
	
	\subsection{Approach}
	
	Consider the symmetric matrix-vector multiplication $$y[i] = A[i, j] * x[j].$$ Because the vector output is non-symmetric, we must compute for all possible $[i, j]$, but because $A$ is symmetric, we can only access the canonical half of $A$.
	
\begin{figure}[t]
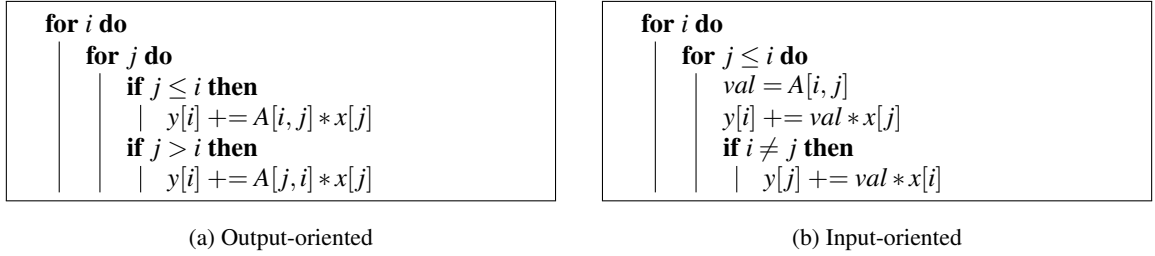

		\vspace{-\baselineskip}
		\centering
		\begin{subfigure}[b]{0.48\textwidth}
			\begin{algorithm}[H]
				\For{$i$} {
					\For{$j$} {
						\uIf{$j \leq i$} {
							$y[i] \mathrel{+}= A[i, j] * x[j]$
						}
        						\uIf{$j > i$} {
            						$y[i] \mathrel{+}= A[j, i] * x[j]$
						}
					}
				}
			\end{algorithm}
			\caption{Output-oriented}
			\label{loops2:a}
		\end{subfigure}
		\hfill
		\begin{subfigure}[b]{0.48\textwidth}
			\begin{algorithm}[H]
				\For{$i$} {
					\For{$j \leq i$} {
						$val = A[i, j]$\\
						$y[i] \mathrel{+}= val * x[j]$\\
						\uIf{$i \neq j$} {
							$y[j] \mathrel{+}= val * x[i]$
						}					
					}
        				}
			\end{algorithm}
			\caption{Input-oriented}
			\label{loops2:b}
		\end{subfigure}
		\caption{Two approaches to symmetric matrix-vector multiplication.}
	\end{figure}

	One approach (Figure \ref{loops2:a}) is to iterate through all $[i, j]$, splitting the for-loop into cases where the coordinates are canonical or non-canonical. For the latter, we permute the coordinates to obtain the canonical counterpart. Another approach (Figure \ref{loops2:b}) is to iterate through canonical $[i, j]$. For each canonical matrix access, we write into the output twice: once for the usual computation, and once to perform the computation for the non-canonical counterpart, given by $y[j] = A[j, i] * x[i]$. 
	
	Focus for a moment on the overall loop structure: [for $i$; for $j$] and [for $i$; for $j \leq i$], respectively. We call the first approach \textit{output-oriented} because we iterate through all coordinates needed to sequentially compute the output. We call the second approach \textit{input-oriented} because we iterate through only the coordinates needed to sequentially access the symmetric input(s).
	
	A few differences arise. Both approaches require random access, but output-oriented iteration randomly accesses the symmetric input, while input-oriented iteration randomly accesses the output and non-symmetric inputs. This could affect performance and the (in)ability to allow for some sparsity. For example, Figure \ref{loops2:b} is better suited to the case where the input $A$ is sparse, since it accesses $A$ sequentially. In fact, the algorithm in this figure is very similar to the one used for sparse symmetric matrix-vector multiplication in \cite{de2003performance}.
			
	Observe that there is always only one output, but there can be any number of arbitrarily misaligned symmetric inputs: this makes the first approach more obviously generalizable. As our next example, consider $$C[i, \ell] = A[i, j, k] * B[j, k, \ell],$$ where $A$ has $\{\{i, j\}, \{k\}\}$ symmetry and $B$ has $\{\{j, k\}, \{\ell\}\}$ symmetry. 
	
	If we try to generate code using the input-oriented approach, we immediately encounter a problem: the canonical constraint is $i \geq j$ for $A$ and $j \geq k$ for $B$. We might naively hope to combine these into the requirement that $i \geq j \geq k$, but this will lead to many coordinates being incorrectly skipped --- for $A$, for instance, we need to access all possible $k$. Put another way, we cannot have a top-level loop structure (i.e., without splitting into different cases) that iterates through exclusively the canonical regions of both symmetric inputs. 
	
	A further complication is that in the input-oriented approach, we need to include a series of if-statements to prevent double counting in the case where two or more coordinates have the same value. This gets increasingly more complex for higher dimensions. By contrast, in the output-oriented approach, we always write to the output exactly once per iteration, so this issue is easily resolved by making some inequalities strict.	
	
	In the next section, we focus on applying the output-oriented approach for this and other examples. 
	
	\subsection{Code Generation}

\begin{figure}[t]
	\vspace{-\baselineskip}	
		\centering
		\begin{subfigure}[b]{0.48\textwidth}
			\begin{subfigure}[b]{\textwidth}
			\begin{algorithm}[H]
				\For{$i$} {
					\For{$j \leq i$} {
						\hl{$C[i, j] = A[i, j] + B[i, j]$}
					}
					\For{$j > i$} {
						{\color{lightgray} $C[i, j] = $} $A[j, i]${\color{lightgray} $\,+\,B[i, j]$}
					}
				}
			\end{algorithm}
			\caption{$A$: $\{\{i, j\}\}$ symmetry, $B$: non-symmetric}
			\label{loops3:a}
			\end{subfigure}
			\ \\			
			\begin{subfigure}[b]{\textwidth}
			\begin{algorithm}[H]
				\For{$i$} {
					\For{$j \leq i$} {
						\For{$k \leq j$} {
							\hl{$C[i, j, k] = A[i, j, k] + B[i, j, k]$}
						}
						
						\For{$j < k \leq i$} {
							{\color{lightgray} $C[i, j, k] = $} $A[i, k, j]${\color{lightgray}$\,+\,B[i, j, k]$}
						}
						\For{$k > i$} {
							{\color{lightgray} $C[i, j, k] = $} $A[k, i, j]${\color{lightgray}$\,+\,B[i, j, k]$}
						}
					}
					
					\For{$j > i$} {
						\For{$k \leq i$} {
							{\color{lightgray} $C[i, j, k] = $} $A[j, i, k]${\color{lightgray}$\,+\,B[i, j, k]$}
						}
						
						\For{$i < k \leq j$} {
							{\color{lightgray} $C[i, j, k] = $} $A[j, k, i]${\color{lightgray}$\,+\,B[i, j, k]$}
						}
						\For{$k > j$} {
							{\color{lightgray} $C[i, j, k] = $} $A[k, j, i]${\color{lightgray}$\,+\,B[i, j, k]$}
						}
					}
				}
			\end{algorithm}
			\caption{$A$: $\{\{i, j, k\}\}$ symmetry, $B$: non-symmetric}
			\label{loops3:b}
			\end{subfigure}
		\end{subfigure}
		\hfill 
		\begin{subfigure}[b]{0.48\textwidth}
		\begin{subfigure}[b]{\textwidth}
			\begin{algorithm}[H]
				\For{$i$} {
					\For{$j \leq i$} {
						\For{$k \leq j$} {
							\hl{$C[i, j, k] = A[i, j, k] + B[i, j, k]$}
						}
						
						\For{$j < k \leq i$} {
							{\color{lightgray} $C[i, j, k] = $} $A[i, k, j]${\color{lightgray}$\,+\,B[i, j, k]$}
						}
						\For{$k > i$} {
							{\color{lightgray} $C[i, j, k] = $} $A[k, i, j]${\color{lightgray}$\,+\,B[i, j, k]$}
						}
					}
				}
			\end{algorithm}
			\caption{$A$: $\{\{i, j, k\}\}$ symmetry, $B$: $\{\{i, j\}, \{k\}\}$ symmetry}
			\label{loops3:c}
		\end{subfigure}
		\ \\
		\begin{subfigure}[b]{\textwidth}
			\begin{algorithm}[H]
				\For{$i$} {
					\For{$j \leq i$} {
							\For{$k \leq j$} {
								\For{$\ell$} {
									\hl{$C[i, \ell] \mathrel{+}= A[i, j, k] * B[j, k, \ell]$}
								}
							}
							\For{$k > j$} {
								\For{$\ell$} {
									{\color{lightgray} $C[i, \ell] \mathrel{+}= A[i, j, k]\,*\,$}$B[k, j, \ell]$
								}
							}
					}
					
					\For{$j > i$} {
							\For{$k \leq j$} {
								\For{$\ell$} {								
									{\color{lightgray} $C[i, \ell] \mathrel{+}=$} $A[j, i, k]${\color{lightgray}$\,*\,B[j, k, \ell]$}
								}
							}
							\For{$k > j$} {
								\For{$\ell$} {			
									{\color{lightgray}$C[i, \ell] \mathrel{+}=$} $A[j, i, k]{\color{lightgray}\,*\,}B[k, j, \ell]$
								}
							}
					}
				}
			\end{algorithm}
			\caption{$A$: $\{\{i, j\}, \{k\}\}$ symmetry, $B$: $\{\{j, k\}, \{\ell\}\}$ symmetry}
			\label{loops3:d}
		\end{subfigure}
		\end{subfigure}
		\caption{Examples of code generated using the output-oriented approach. The original computation is highlighted, the unchanged terms are grayed out, and the input symmetries are described in the captions.} 
		\label{loops3}
\end{figure}

	We separate the five main phases into the subsections below. Roughly, we start by defining a series of intermediate representations (Sections 5.2.1, 5.2.2, and 5.2.3) that enables us to determine which for-loops we need, and then we translate these representations into actual code (Sections 5.2.4 and 5.2.5).
	
	\subsubsection{Compute the greatest common symmetry}
	
	Our purpose in defining computational symmetry was to use it to guide our loop structure, so naturally we begin by determining the gcs. Many algorithms are plausible; here is one:
	
	\begin{enumerate}
		\item[A.] Write down all the indices used in the computation, and put them in the same part of a symmetry $S_G$.
		\item[B.] Iterate through each tensor symmetry $S_T$. For the input tensors, just use the provided input symmetry; for the output tensor, temporarily assume that it is fully symmetric.
		\item[C.] Set $S_G \leftarrow S_G / S_T$.
		That is, for each part $p_g$ of $S_G$ and each part $p_t$ of $S_T$, split $p_g$ into two parts, the intersection $p_g \cap p_t$ and the remaining indices $p_g - p_t$. (Empty parts can be discarded.)
		
		For example, if $S_G = \{\{i, j, k, \ell\}, \{m, n\}\}$ and $S_T = \{\{k, \ell, m, n\}\}$, then $S_G / S_T = \{\{i, j\}, \{k, \ell\}, \{m, n\}\}$.
		
	\end{enumerate}
		
	\noindent To further visualize this procedure, it may help to revisit Figure \ref{boxes}.
	
	\subsubsection{Record dependencies} 
		
	Since we define coordinates to be canonical if they respect a specific ordering relative to the tensor's symmetry, we will use an \textit{index dependency graph} to indicate which indices' orderings matter in the context of our computation. For example, Figure \ref{loops3:a}, which slightly rephrases the earlier Figure \ref{loops2:a}, has the graph below, where the directed edge from the node for $j$ to the node for $i$ means that we need to consider the coordinate value of $j$ relative to $i$. We also say $i$ is $j$'s parent.
	
	\begin{center}
	\begin{tikzpicture}[main/.style = {draw, circle, inner sep=0pt, minimum width = 0.75cm}] 
		\node[main] (1) {$i$}; 
		\node[main] (2) [right = 0.5 cm of 1] {$j$}; 
		\draw[->] (2) -- (1);
	\end{tikzpicture}
	\end{center}
	
	Suppose we have a computation using $n$ index variables labeled such that the iteration order is $i_1$, \dots, $i_n$.  We construct index dependency graphs by incorporating information from the input symmetries:
	
	\begin{enumerate}
		\item[A.] Start with a node for each index variable and no edges.
		
		\item[B.] The parts of a symmetry partition are sets, but we can make them into lists sorted by their place in the iteration order. In this way, the part $\{i_1, i_2\}$ or, equivalently, $\{i_2, i_1\}$ becomes $[i_1, i_2]$. 
		
		For each part in each input symmetry, attempt to add an edge between each variable $i_m$ and the variable that immediately precedes it in the list. 
		
		\item[C.] If there is an already an edge from $i_m$ to some other variable, then let the variable that is later in the iteration order be $i_\ell$ and the variable that is earlier be $i_e$. Then, replace the current edge with an edge from $i_m$ to $i_\ell$, and add an edge from the least ancestor of $i_\ell$ (possibly itself) to $i_e$. 
		
	\end{enumerate} 

	\noindent We elaborate through the example below.
	
	\begin{example}
		Consider three input tensors that use index variables $i$, $j$, $k$, $\ell$, $m$, in that iteration order.
		\begin{center}
			\begin{tikzpicture}[main/.style = {draw, circle, inner sep=0pt, minimum width = 0.75cm}] 
				\node[main] (1) {$i$}; 
				\node[main] (2) [right = 0.5 cm of 1] {$j$}; 
				\node[main] (3) [right = 0.5 cm of 2] {$k$}; 
				\node[main] (4) [right = 0.5 cm of 3] {$\ell$}; 
				\node[main] (5) [right = 0.5 cm of 4] {$m$}; 
			\end{tikzpicture}
		\end{center}
		Suppose the first tensor has symmetry $\{\{i, \ell\}, \{j, k\}\}$. Then we add one edge for each part.
		\begin{center}
			\begin{tikzpicture}[main/.style = {draw, circle, inner sep=0pt, minimum width = 0.75cm}] 
				\node[main] (1) {$i$}; 
				\node[main] (2) [right = 0.5 cm of 1] {$j$}; 
				\node[main] (3) [right = 0.5 cm of 2] {$k$}; 
				\node[main] (4) [right = 0.5 cm of 3] {$\ell$}; 
				\node[main] (5) [right = 0.5 cm of 4] {$m$}; 
				\draw[->] (4) to [out=150,in=30] (1); 
				\draw[->] (3) -- (2);
			\end{tikzpicture}
		\end{center}
		Next, suppose the second tensor has symmetry $\{\{i, m\}\}$. Another edge is added.
		\begin{center}
			\begin{tikzpicture}[main/.style = {draw, circle, inner sep=0pt, minimum width = 0.75cm}] 
				\node[main] (1) {$i$}; 
				\node[main] (2) [right = 0.5 cm of 1] {$j$}; 
				\node[main] (3) [right = 0.5 cm of 2] {$k$}; 
				\node[main] (4) [right = 0.5 cm of 3] {$\ell$}; 
				\node[main] (5) [right = 0.5 cm of 4] {$m$}; 
				\draw[->] (4) to [out=150,in=30] (1); 
				\draw[->] (3) -- (2);
				\draw[->] (5) to [out=-150,in=-30] (1); 
			\end{tikzpicture}
		\end{center}
		Suppose the final tensor has symmetry $\{\{k, \ell\}\}$. Here, Step C comes into play. We want to add an edge from $\ell$ to $k$, but $\ell$ already has an edge to $i$. Since $k$ comes later in the iteration order, we have $i_\ell = k$ and $i_e = i$ in the previous notation. Hence we now have that $k$ is $\ell$'s parent, and $i$ is $j$'s parent. 
		
		\begin{center}
			\begin{tikzpicture}[main/.style = {draw, circle, inner sep=0pt, minimum width = 0.75cm}] 
				\node[main] (1) {$i$}; 
				\node[main] (2) [right = 0.5 cm of 1] {$j$}; 
				\node[main] (3) [right = 0.5 cm of 2] {$k$}; 
				\node[main] (4) [right = 0.5 cm of 3] {$\ell$}; 
				\node[main] (5) [right = 0.5 cm of 4] {$m$}; 
				\draw[->] (4) -- (3); 
				\draw[->] (3) -- (2);
				\draw[->] (2) -- (1);
				\draw[->] (5) to [out=-150,in=-30] (1); 
			\end{tikzpicture}
		\end{center}
		Observe that two variables can have the same parent, but a variable may not have two parents.
	\end{example}

	\subsubsection{Enumerate orderings}
	
	Equipped with the greatest common symmetry and the index dependency graph, we can now build one last intermediate representation. We will use an \textit{ordering tree} to enumerate the possible index orderings and thus make explicit which for-loops should appear in the code. Returning to the example in Figure \ref{loops3:a}, we have
	\begin{align*}
		&\text{[}{\color{magenta} i}\text{]}\\
		&\null \quad \text{[}{\color{teal} j}, i\text{]}\\
		&\null \quad \text{[}i, {\color{teal} j}\text{]}
	\end{align*}
	The tree above says that the outermost loop is simply [for $i$]. Nested inside (as indicated by the indentation) are two additional loops. Each ordering represents a subdomain of coordinates, constrained to be non-decreasing. So, [{\color{teal} $j$}, $i$] corresponds to the loop [for $j \leq i$] and [$i$, {\color{teal} $j$}] corresponds to the loop [for $j \geq i$]. The colored index is the one that we are iterating over.
	
	We describe the general algorithm for $i_1, \dots, i_n$ on the left and an example on the right for this graph. 	\begin{center}
		\begin{tikzpicture}[main/.style = {draw, circle, inner sep=0pt, minimum width = 0.75cm}] 
			\node[main] (1) {$i$}; 
			\node[main] (2) [right = 0.5 cm of 1] {$j$}; 
			\node[main] (3) [right = 0.5 cm of 2] {$k$}; 
			
			\draw[->] (2) -- (1); 
			\draw[->] (3) -- (2); 
		\end{tikzpicture}
	\end{center}
	
	\noindent\begin{minipage}[t]{0.65\textwidth}
	\begin{enumerate}
		\item[A.] Start with the loop for the first variable $i_1$. (Line 1.)
		\item[B.] With each subsequent $i_m$, insert it in all possible positions relative to the orderings for its parent in the graph. 
		
		(Lines 2 and 6 for $j$. Lines 3-5 and 7-9 for $k$.)		
		\item[C.] Skip orderings that are non-canonical relative to the gcs.
		
		For example, if the gcs is $\{\{i, j\}, \{k\}\}$, we discard orderings that place $i$ before $j$, thereby omitting Lines 6-9. 
	\end{enumerate}
	\end{minipage}
	\hfill 
	\begin{minipage}[t]{0.22\textwidth}
		1 \quad [{\color{magenta}$i$}]\\
		2 \quad \quad [{\color{teal} $j$}, $i$]\\
		3 \quad \quad \quad [{\color{orange} $k$}, $j$, $i$]\\
		4 \quad \quad \quad [$j$, {\color{orange} $k$}, $i$]\\
		5 \quad \quad \quad [$j$, $i$, {\color{orange} $k$}]\\
		6 \quad \quad [$i$, {\color{teal} $j$}]\\
		7 \quad \quad \quad [{\color{orange} $k$}, $i$, $j$]\\
		8 \quad \quad \quad [$i$, {\color{orange} $k$}, $j$]\\
		9 \quad \quad \quad [$i$, $j$, {\color{orange} $k$}]\\
	\end{minipage}		
	
	\subsubsection{Generate loops}
	The next set of steps is to convert these ordering trees into the actual for-loops and statements. Lines 1-9 will eventually correspond to Figure \ref{loops3:b}, while Lines 1-5 will eventually correspond to Figure \ref{loops3:c}.
	
	\begin{enumerate}
		\item[A.] Given an ordering for some $i_m$, if the preceding variable is $i_p$ and the succeeding variable is $i_s$, the loop is [for $i_p < i_m \leq i_s$]. The lower bound is 0 when there is no predecessor; the upper bound is the size of that tensor dimension when there is no successor.
				
		\item[B.] Within the innermost for-loop, modify the original statement such that the accesses are canonical. For example, for $j < k \leq i$ in the context of Figure \ref{loops3:b}, we need to permute the coordinates of $A$, obtaining $$C[i, j, k] = A[i, k, j] + B[i, j, k].$$
	\end{enumerate}
	
	A problem with our formulation of index dependency graphs is that we sometimes encode unnecessary orderings, leading to redundant loops. Take Figure \ref{loops3:d}, for example. Due to the $\{\{i, j\}\}$ part in $A$ and the $\{\{j, k\}\}$ part in $B$, the graph will have edges from $k$ to $j$ and from $j$ to $i$. Hence we transitively consider all possible orderings of $i$, $j$, and $k$, but two loops that differ only by the relative coordinate values of $i$ and $k$ will have the same accesses. So we will generate both of the following loops for $j \leq i$ (and likewise for $j > i$): 
	\begin{center}
	\RestyleAlgo{plain}
		\begin{minipage}{0.4\textwidth}
		\begin{algorithm}[H]
			\For{$j < k \leq i$} {
				\For{$\ell$} {
					$C[i, \ell] \mathrel{+}= A[i, j, k] * B[k, j, \ell]$
				}
		}
		\end{algorithm}
		\end{minipage}
		\ \
		\begin{minipage}{0.4\textwidth}		
		\begin{algorithm}[H]
			\For{$k > i$} {
				\For{$\ell$} {
					$C[i, \ell] \mathrel{+}= A[i, j, k] * B[k, j, \ell]$
				}
			}
		\end{algorithm}
		\end{minipage}
	\end{center}
	The process of eliminating these redundancies is simple: merge the bounds of adjacent loops whose inner contents are the same. In this case, the new loop has bounds $k > j$. The final result is shown in Figure \ref{loops3:d}.
	
	However, the presence of redundant loops indicates that the graph does not capture as much information as it should. Once we add an edge to the graph, we do not mark which input tensor contributed the edge. This prevents us from distinguishing between the case where one tensor, say, has $\{\{i, j, k\}\}$ symmetry, which would indeed require all possible orderings, and this case, where two separate tensors have $\{\{i, j\}\}$ and $\{\{j, k\}\}$ symmetry. A more complex notion of index dependency graphs may alleviate this problem.
	
	\subsubsection{Lower the concrete index notation}
	
	\begin{figure}		
	\begin{subfigure}[b]{\textwidth}
		\vspace{-\baselineskip}	
		\includegraphics[width=\textwidth]{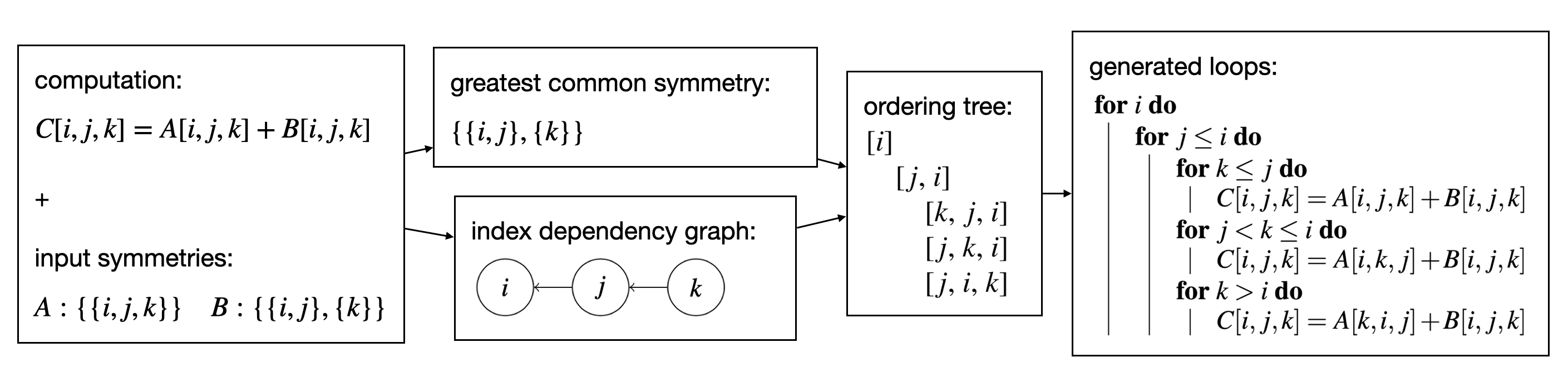}
	\end{subfigure}
	\ \\
	\begin{subfigure}[b]{\textwidth}
		\vspace{-\baselineskip}	
		\includegraphics[width=\textwidth]{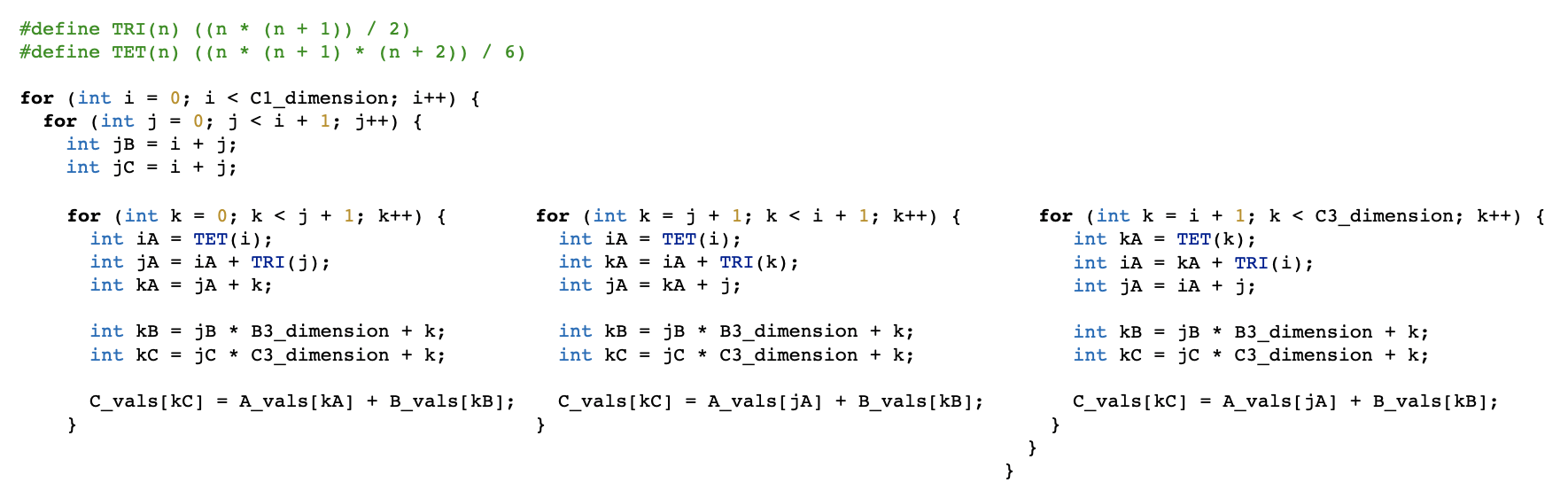}
	\end{subfigure}
	
	\caption{An illustration of the step-by-step process for generating the loop structure and the resultant C code.}
	\label{bigcode}
	\end{figure}
	
	An abuse of terminology that we have been employing is that the ``code'' we have generated thus far is in fact styled after the \textit{concrete index notation} used in taco \cite{kjolstad2017tensor}. We do so because we consider the implementation details of taco largely as a black box, useful for steps like representing the tensor expressions through various data types, determining the iteration order and performing concretization steps not directly related to symmetry, and finally, lowering the loops into actual code.
	
	A few extensions to taco are needed to accommodate symmetry. The triangular indexing scheme in Section \ref{tri} may be added as a new tensor format, while the idea of separating loops into different cases may be supported through variable for-loop bounds and representing the loops for each index variable as a multi statement of forall statements.
	
	The example in Figure \ref{loops3:c} is illustrated in full in Figure \ref{bigcode}.
	\subsection{Code Evaluation}
	
	The primary motivation in developing this approach was to ensure that we can generate code for any computation (expressible in our notation), with any input symmetries (expressible through our definition), in a way that takes full advantage of storage savings. 
	
	However, in the process, we incur significant hits in performance when the symmetries are misaligned, likely due to the way we randomly access certain inputs. Due to time constraints, benchmarking data was collected informally and thus cannot be displayed here, but it appears that even for simple cases such as three-dimensional addition of a fully symmetric and a non-symmetric tensor, our code performs considerably worse than the original taco code. 
	
	\section{Related Works}
	
	Prior works that examine symmetric tensor computation can be roughly grouped in two categories:
	
	First, a few library-based approaches exist, especially in the field of computational chemistry. For example, libtensor \cite{epifanovsky2013new} maintains symmetric tensors in a block structure, where ``canonical'' blocks store the actual values, while ``derivative'' blocks point to their canonical counterparts. The cyclops tensor framework (CTF) \cite{solomonik2012preliminary}, on the other hand, uses cyclic decomposition to aid parallelization. 
		
	Our work is closer to libtensor than to CTF, but still differs in a few main ways: we apply a compiler-based approach, and we explore the design implications of using an unblocked storage scheme without pointers. One limitation is that we do not support as many types of symmetry; for example, we do not address point-group or spin symmetry, as they appear to be more specific to physics or chemistry. 
	
	The other category consists of papers that are largely based around a particular operation. One paper \cite{schatz2014exploiting} centers on ``symmetric tensor times same matrix,'' which can be thought of as a generalization of a change-of-basis. We use the definition of partial symmetry provided in this paper but otherwise diverge significantly: while they leverage the details of this particular operation, our broader focus leads to questions regarding how to compute with arbitrary tensors possessing arbitrary symmetries.
	
	\section{Conclusion} 
	
	Our attempt to develop a compiler-based approach for symmetric tensor computations can perhaps be summarized by this chief difficulty. Symmetric storage is highly personalized to each tensor's symmetry: in our case, we store only the canonical coordinates. As a result, computing with tensors that have misaligned symmetries adds significant complexity. 
	
	We focus on resolving this complexity over the course of code generation, so that the produced loop structure is simple. However, in doing so, we did not meet our goal of also reducing computation time. This suggests that to generate efficient code, we need major optimizations or a different approach altogether.
		
	\nocite{*}

	\newpage
	\bibliographystyle{plain}
	\bibliography{refs}

\end{document}